# Metallic iron nanocubes for magnetic hyperthermia :
# Large specific absorption rates and ways towards further improvements.

B. Mehdaoui, A. Meffre, L.-M. Lacroix, J. Carrey*, S. Lachaize, M. Respaud

Université de Toulouse; INSA; UPS; LPCNO (Laboratoire de Physique et Chimie des Nano-Objets), 135 avenue de Rangueil, F-31077 Toulouse, France and
CNRS; UMR 5215 ; LPCNO, F-31077 Toulouse, France

M. Gougeon

Institut CARNOT - CIRIMAT - UMR 5085, Bâtiment 2R1, 118 route de Narbonne
F-31062 Toulouse, France

B. Chaudret

Laboratoire de Chimie de Coordination-CNRS, 205 rte de Narbonne, 31077 Toulouse cedex 4, France

**Abstract:**
We report the magnetic hyperthermia properties of chemically synthesized ferromagnetic 11 and 16 nm Fe nanocubes displaying the saturation magnetization of bulk iron. The specific absorption rate measured on 16 nm nanocubes is 1690±160 W/g at 300 kHz and 66 mT. The influence of magnetic interactions on their magnetic properties and on their efficiency for hyperthermia is discussed. Based on the Stoner-Wohlfarth model, conditions on their diameter and on their coating are derived to further increase their efficiency.

**Main Text:**

Magnetic hyperthermia is a way to improve the efficiency of chemotherapy or radiotherapy by raising the temperature of a tumour to 41-45 °C during a few hours using magnetic nanoparticles (MNPs): in the clinical treatment set at the Charité Hospital, Berlin, MNPs are first localised inside the tumour and then excited by an alternating magnetic field of moderate amplitude $\mu_0 H_{app}$ (12-25 mT) at a frequency $f_{exc}$ of 100 kHz [1]. The power released by the NPs is assessed by their specific absorption rate (SAR) or their specific losses $A$, linked by the equation SAR = $Af_{exc}$. Increasing SARs above 1 kW/g could be beneficial for several aspects of the hyperthermia applications but it constitutes a real challenge [2]. Indeed, due to physiological issues, human body cannot be exposed to alternating magnetic field presenting a large $H_{app}f_{exc}$ [2]. Therefore, the optimization must only rely on the MNPs. The most favourable case, leading to $SAR_{max} = 4\mu_0 \sigma_S H_{app} f_{exc}$ occurs when i) the magnetization loop, characterized by a coercive field $H_C$ and a saturation magnetization per unit mass $\sigma_S$, is a perfect square and ii) $H_C \approx H_{app}$. Unfortunately, due to random orientations of easy axis, to magnetic interactions and to thermal activation, hysteresis loops of MNPs assemblies are far from being square. Thus, more generally, one can write:

$$SAR = 4\alpha\mu_0 \sigma_S H_{app} f_{exc} \qquad (1)$$

where $\alpha$ is a dimensionless parameter which characterizes the relative area of the hysteresis loops with respect to the ideal square. As a consequence of Equ. (1) and of the limitation of the $H_{app}f_{exc}$ product, the maximisation of the SAR requires MNPs displaying a high $\sigma_S$ and hysteresis loops as square as possible (large α).

High magnetization materials are essentially metallic Ni, Fe, Co and their alloys. Among them, iron is the most promising one since it displays a very large $\sigma_S$ and is *a priori* the most biocompatible. However the most widely studied materials for magnetic hyperthermia have been, so far, the iron oxides because of their full biocompatibility and the relative simplicity of their synthesis and of their handling. Optimized chemically-synthesized $Fe_3O_4$ NPs have shown losses up to 1.5 mJ/g at 400 kHz [3]. Among high magnetization materials, higher loss values up to 3.2 mJ/g at 400 kHz have been reported for Co MNPs [4]. The scarce results published so far on Fe MNPs are disappointing as a consequence of the lack of control of the surface oxidation: hyperthermia experiments were performed on Fe/$Fe_xO_y$ core-shell MNPs which displayed low $\sigma_S$ and losses comparable to those measured on iron oxides [5].

Our group has developed an organometallic synthesis allowing the controlled synthesis of pure metallic iron NPs displaying the bulk magnetization [6,7,8]. We report in this article the magnetic and hyperthermia measurements on ferromagnetic 11 and 16 nm nanocubes displaying very large losses. These results are discussed in the aim to define the way toward an optimized performance, based on the size-dependence of the coercive field in single-domain MNPs.

The two samples of Fe nanocubes were synthesized by a two steps organometallic route which has been previously described [8]. A complete description of the size and shape control on iron NPs synthesized by this route will be published elsewhere. First, we prepared small Fe seeds (~2 nm) through the decomposition at 150°C for 24h of a mesitylene (20mL) solution of the iron dimer {Fe[N(SiMe$_3$)$_2$]$_2$}$_2$, (376 mg, 0.5 mmol) under 3 bars of $H_2$ in a Fisher-Porter bottle (170mL) [9]. In a second time, a mixture of hexadecylammonium chloride (HDA.HCL) and hexadecylamine (HDA) was added to the colloidal solution. For sample 1, the ratio ammonium/amine was 1:2 (277 mg, 1 mmol HDA.HCL, 483 mg, 2 mmols HDA). For sample 2, this ratio was 1.1:2 (304.7 mg, 1.1 mmol HDA.HCL, 483 mg, 2 mmols HDA). The solution was stirred for 20 minutes at 90°C then pressurised under 3 bars of $H_2$ and heated at 150°C. After 48h, a black precipitate was formed at the bottom of the Fisher-Porter bottle. The solvent was filtered off and the precipitate washed three times with 15 mL of toluene to remove the surfactants in excess and other remaining molecular species. The final iron content was 71 % and 88 % for samples 1 and 2 respectively, as determined by chemical analysis.

Samples for transmission electron microscopy (TEM) were prepared by the deposition of a drop of solution onto a carbon-coated copper grid. High-resolution TEM (not shown) revealed that the nanocubes were single-crystalline and exhibited a bcc crystal structure, with facets of (100) planes. As revealed by TEM micrographs, most of the nanocubes were embedded in organic mesophases, which prevent them to form a true colloidal solution (see Fig. 1). Similar nanocube-filled mesophases have already been characterized for a similar system in which carboxylic acid was used instead of ammonium chloride [6]. Size distributions measured from several TEM micrographs showed that samples 1 and 2 were composed of iron nanocubes of mean side lengths 16.3 ± 1.5 nm (sample 1) and 11.3 ± 1.3 nm (sample 2) respectively.

Samples for magnetic measurements were prepared and sealed under an argon atmosphere in order to preserve the metallic character of Fe. SQUID measurements on powders of samples 1 and 2 are shown in Fig. 2. Their saturation magnetizations per unit mass $\sigma_S$ were 200±10 $Am^2/kg$ and 178±9 $Am^2/kg$ at 300 K, respectively, just below the bulk value. Their coercive fields $H_C$ at 300 K are 16 and 5 mT, respectively.

Hyperthermia experiments were performed on an induction oven working at a frequency of 300 kHz and a maximum magnetic field of 66 mT. For hyperthermia measurements, an ampoule containing the colloidal solution was sealed under vacuum to prevent any oxidation of their NPs. A typical ampoule contains 9 mg of powder and 550 mg of mesitylene. The ampoule was then placed into a calorimeter with 1.5 mL of deionised water, the temperature of which was measured. The measurement time was varied between 20 and 200 seconds, depending on the experiment, so that the temperature rise never exceeds 20°C. After the magnetic field stops, the water is shaken during roughly 20s to ensure the ampoule thermalization and the homogeneity of the water temperature, which is checked by putting two probes at the top and the bottom of the calorimeter. The temperature rise is measured after this process. For measurements longer than 20s, the raw SAR values were corrected from the calorimeter losses, which were previously calibrated. The SAR values at 66 mT and their error bars were obtained by averaging three measurements of 20s on three different ampoules arising from the same synthesis (9 values). The complete magnetic-field dependence of the SAR was measured on a single ampoule for each synthesis and its SAR value renormalized accordingly.

During hyperthermia experiments, the spatial organisation of the MNPs in the ampoule was completely redistributed by the application of the magnetic field. Indeed, while the MNPs aggregates fall down by gravity at the bottom of the ampoule in the absence of any applied magnetic field, the MNPs formed small spikes along the field direction when a small magnetic field was applied. Moreover, for larger magnetic fields, the MNPs self-organized into regularly spaced levitating needles, which disintegrated as soon as the magnetic field was stopped. A video illustrating this phenomenon in sample 1 is shown in Fig. 3(a). The magnetic field for which these needles formed was higher for sample 1 than for sample 2 and ranged between 20 to 30 mT. The formation of columns in a magnetic field is a classical behaviour of ferrofluids and is due to magnetic interactions between the NPs [10].

Fig. 3(b) displays the SAR values at 300 kHz as a function of the applied magnetic field for the two samples. In both cases, SAR increases strongly in the range 10-30 mT and then follows a roughly linear increase at higher magnetic field without complete saturation. For sample 1, the increase is sharper and occurs at a higher magnetic field than for sample 2. Such an abrupt increase followed by a plateau is a typical feature of ferromagnetic samples and was previously reported on ferromagnetic FeCo MNPs [11] : the sharp rise of the SAR occurs when $H_{app}$ reaches the coercive field of the MNPs. SARs up to 1690±160 W/g and 1320±140 W/g are measured at 66 mT for samples 1 and 2, respectively. For sample 1, this value corresponds to losses of 5.6±0.5 mJ/g. These losses exceed by a factor 3 the values reported on optimized chemically-synthesized iron oxide NPs [3].

However the efficiency of our system is limited for two reasons. First, the calculation of $\alpha$ at $\mu_0 H_{app} = 66$ mT using Equ. (1) and the experimental $\sigma_S$ value at 300 K leads to 0.11 and 0.09 for samples 1 and 2 respectively. As detailed in the introduction, this parameter gives an estimate of the efficiency for hyperthermia applications of any system. In the case of the Stoner-Wohlfarth (SW) model, $\alpha_{SW}$ is equal to 0.25, considering that i) the MNPs have 3-D randomly distributed anisotropy axis, ii) the MNPs are magnetically independent, iii) the effect of temperature on the magnetization reversal process is neglected, iv) the condition $H_{app} = 2H_C$ leads to the saturation of the NPs and v) the MNPs do not physically rotate in the magnetic field (no Brownian motion) [12]. Moreover, a value of $\alpha = 0.3$ has been measured on randomly oriented optimized iron oxide NPs [3], thus evidencing that reaching $\alpha_{SW}$ is a reasonable experimental expectation. This is not fulfilled in our case. We argue that the presence of magnetic interactions is the main reason for these low $\alpha$ values. Indeed, it is now well admitted that both $H_C$ and $M_R$ decrease in the presence of such interactions [13], thus reducing $\alpha$ below $\alpha_{SW}$. Several observations confirm the presence of magnetic interactions in our samples : i) the nanoparticles form dense needles when the magnetic field is on, ii) the $M_R/M_S$ ratio ($M_R$ is the remnant magnetization and $M_S$ the saturation magnetization) is 0.2 and 0.06 for samples 1 and 2 respectively, whereas 0.5 is expected in the absence of interactions, iii) the saturation field is well above $2H_C$. The second limitation of our system is that we measured huge losses at a magnetic field of 66 mT while, so far, only magnetic fields close to 20 mT were applied during medical hyperthermia treatments [14].

This analysis shows that both the magnetic interactions and the saturation field should be reduced in our system. To cancel the magnetic interactions one needs to work with colloidal solutions in which MNPs are well separated from each other; larger $\alpha$ are then expected. In this case, we recently show that the losses below the blocking temperature of magnetically independent NPs are satisfactorily described by an extended Stoner-Wohlfarth model [11, 15, 16]. In this model, the saturation field is twice the coercive field $\mu_0 H_C$, with:

$$\mu_0 H_C = \frac{2K}{\rho\sigma_S}\left[0.479 - 0.81\left(\frac{k_B T}{2KV}(\ln 1/f_{exc}\tau_0)\right)^{\frac{3}{4}}\right] \qquad (2)$$

where $V$ is the MNP volume, $K$ its anisotropy, $\rho$ its density and $\tau_0 = 10^{-10}$ s. Thus, targeting $\mu_0 H_C$=10 mT for a working magnetic field of 20 mT at 100 kHz would require MNPs of 15 nm if they display the bulk anisotropy. With an enhanced anisotropy compared to the bulk, this optimal size should be reduced accordingly. A last criterion concerns the Brownian motion: Equ. (2) is only valid if the MNPs do not rotate in the magnetic field. This requires that $V > \frac{k_B T}{3\eta f_{exc}}$, where $\eta$ is the viscosity of the medium. This leads to a hydrodynamic NP diameter above 15 nm, using the viscosity of water.

Finally, all these conditions would be fulfilled by core-shell NPs with an iron core of the optimum magnetic volume and a shell thick enough to prevent magnetic interactions and Brownian motion. Such a shell is in any case required for *in-vivo* applications to ensure biodisponibility and targeting of the MNPs. Silica or polyethylene glycol shells [17] can for example increase the hydrodynamic diameter of NPs from 10 to 100 nm. These non-trivial post-treatments will constitute the future developments of this work.

Our measurements of very large SARs on metallic iron NPs above 1 kW/g confirm the potential of high magnetization MNPs for future hyperthermia applications. Size and coating adjustments should be used to increase their efficiency by controlling their coercive field, suppressing the Brownian motion and reducing magnetic interactions. Moreover, the exact optimal size should be tuned as a function on the anisotropy of the synthesized MNPs. These dependencies illustrates that hyperthermia applications still require improvements in the synthesis of colloidal solutions, to produce individual coated NPs with varying size, and need basic research and detailed measurements on their magnetic properties, especially on their anisotropy.

### Acknowledgements:
We acknowledge InNaBioSanté foundation, AO3 program from Université Paul Sabatier (Toulouse) and Conseil Régional de Midi-Pyrénées for financial support, V. Collière (TEMSCAN) for HRTEM, C. Crouzet for technical assistance and A. Mari for magnetic measurements.

# References :

[*] Electronic mail : julian.carrey@insa-toulouse.fr

**Figures**

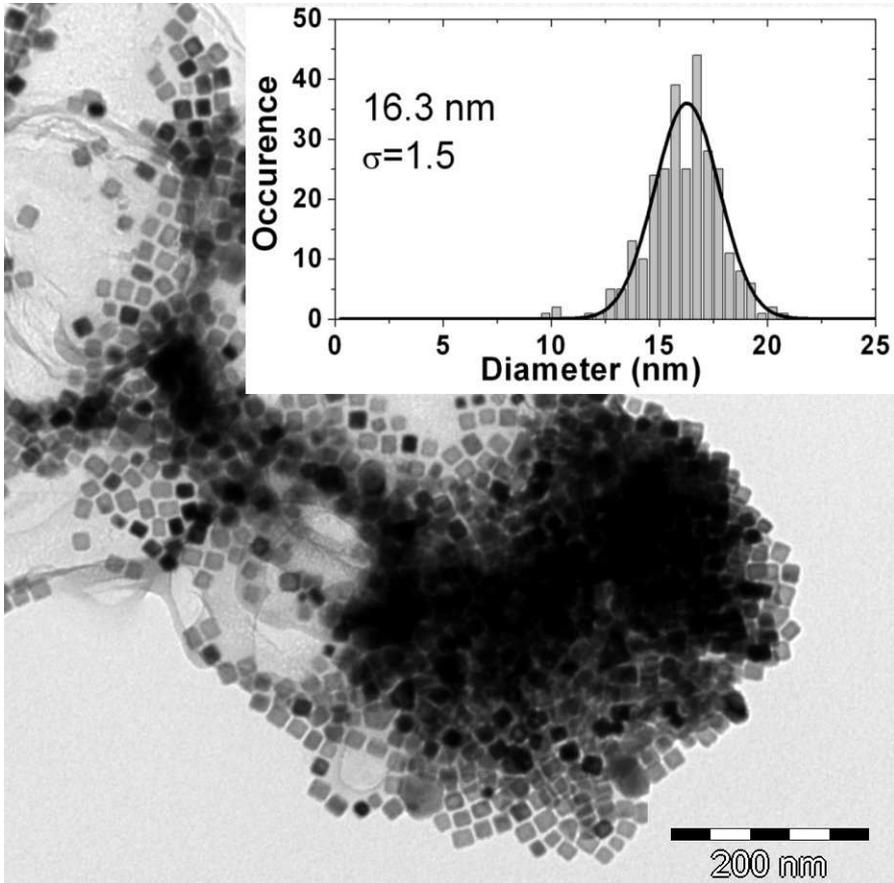

Fig. 1 : TEM micrographs of sample 1. The inset shows the size distribution.

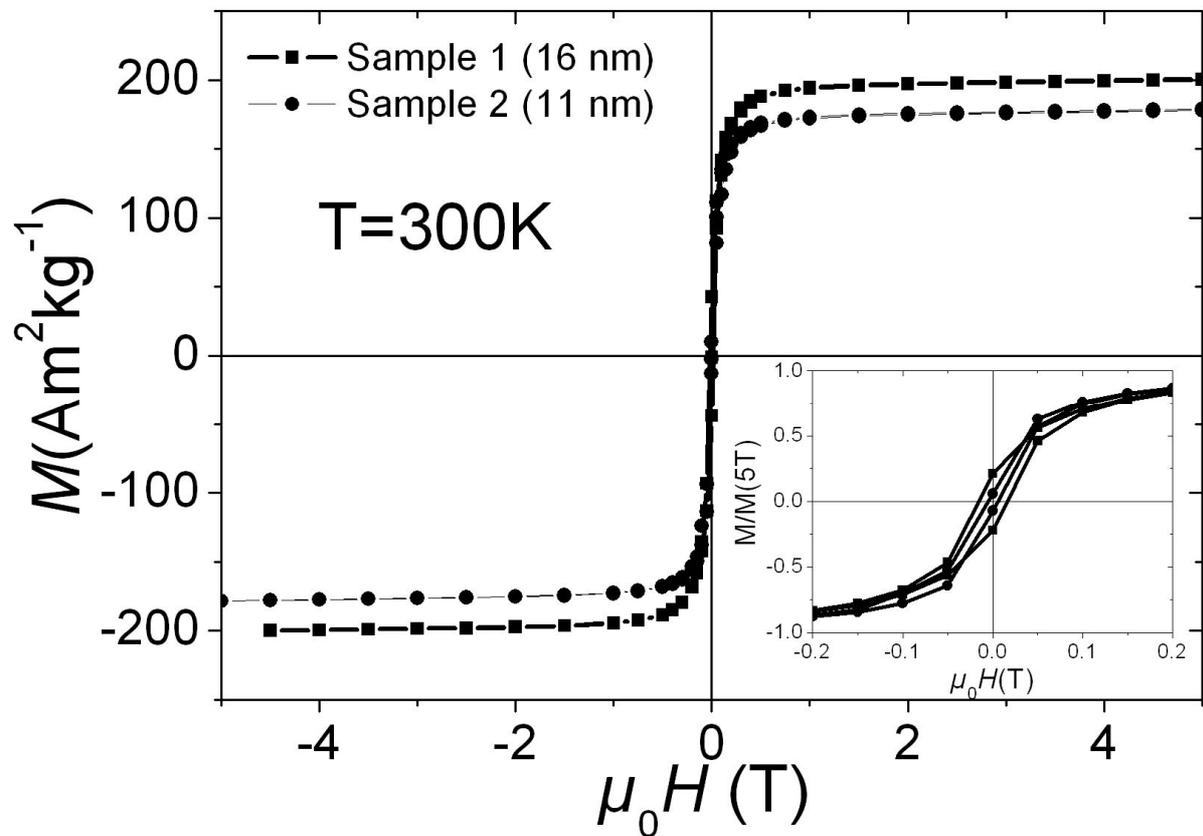

Fig. 2: SQUID measurement at 300 K on sample 1 (16 nm) and sample 2 (11 nm). The inset shows en enlarged view of the magnetization normalized by its value at 5 T.

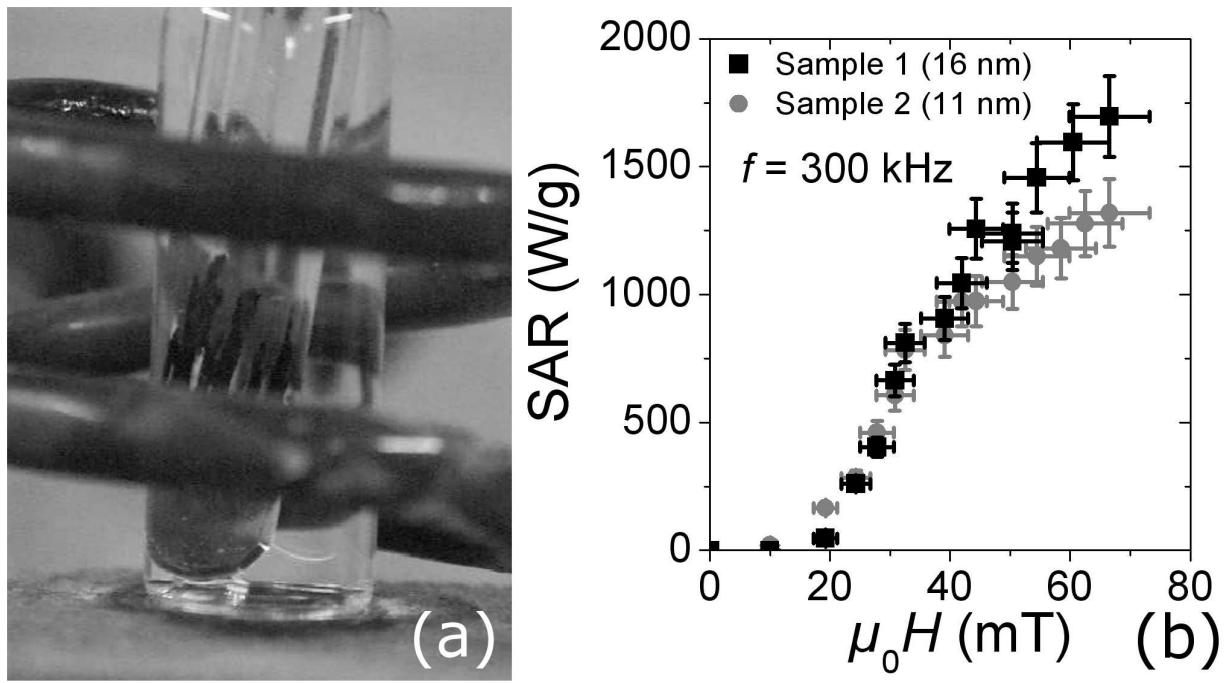

Fig. 3 : (a) Picture extracted from a video showing an hyperthermia experiment on sample 1 at a magnetic field of 66 mT and 300 kHz, when the field is switched on and then off. On this static image, the field is on (enhanced online). (b) Magnetic field dependence of SAR at 300 kHz for the two samples.